# Generative AI Usage and Exam Performance

Janik Ole Wecks [a b] *

ORCID: https://orcid.org/0009-0000-4116-1888
LinkedIn: https://www.linkedin.com/in/janik-ole-wecks-960ba01b2/

Johannes Voshaar [a b]

ORCID: https://orcid.org/0000-0003-0276-4509
LinkedIn: https://www.linkedin.com/in/johannes-voshaar-431aa6201

Benedikt J. Plate [a]

ORCID: https://orcid.org/0009-0003-6982-3718
LinkedIn: https://www.linkedin.com/in/benedikt-j-plate-710a2b144/

Jochen Zimmermann [a]

ORCID: https://orcid.org/0000-0002-1189-7007

[a] Faculty of Business Studies and Economics, University of Bremen, Bremen, Germany
[b] John Molson School of Business, Concordia University, Montreal, Canada

*This version: November 2024*

## Abstract

This study evaluates the impact of students' usage of generative artificial intelligence (GenAI) tools such as ChatGPT on their exam performance. We analyse student essays using GenAI detection systems to identify GenAI users among the cohort. Employing multivariate regression analysis, we find that students using GenAI tools score on average 6.71 (out of 100) points lower than non-users. While GenAI may offer benefits for learning and engagement, the way students actually use it correlates with diminished exam outcomes. Exploring the underlying mechanism, additional analyses show that the effect is particularly detrimental to students with high learning potential, suggesting an effect whereby GenAI tool usage hinders learning. Our findings provide important empirical evidence for the ongoing debate on the integration of GenAI in higher education and underscores the necessity for educators, institutions, and policymakers to carefully consider its implications for student performance.

*Corresponding author. Address: Faculty of Business Studies and Economics, University of Bremen, Max-von-Laue-Str. 1, 28213 Bremen, Germany; Tel.: +49 (421) 218 666 89; Email: wecks@uni-bremen.de.

# 1 Introduction

The launch of OpenAI's user-friendly and conversational ChatGPT in November 2022 made generative artificial intelligence (GenAI) widely accessible to a broad audience, regardless of technical proficiency (Kishore et al., 2023). ChatGPT can process and generate natural language and performs exceptionally in solving real-world problems through back-and-forth conversations, question-answering, and machine translation (Lee, 2023). These novel characteristics led to a tremendous surge in public attention, with over 100 million monthly active users two months after its launch (Ahangama, 2023; Gregor, 2024). The launch spurred heated discussions on the implications of GenAI across various sectors of society. Researchers and media voiced concern about its potential for spreading misinformation, undermining trust (Hsu & Thompson, 2023), and threatening democratic processes and social cohesion (Ferrara, 2024).

The popularity of ChatGPT among students has sparked extensive debate about what role GenAI tools should play in higher education (e.g., Katavic et al., 2023; Strzelecki, 2023). GenAI tools offer considerable benefits such as enabling personalized learning and adaptive instruction, enhancing learning efficiency and student engagement, as well as providing intelligent tutoring systems including real-time feedback, hints, and scaffolding (Kishore et al., 2023). Nevertheless, these tools may hinder students' ability to think independently and critically to solve problems; they also harbour strong potential for perpetuating biases and misinformation (e.g., Kishore et al., 2023). Some educational institutions have thus prohibited the use of ChatGPT at school or blocked it on school devices and networks (e.g., Weber-Wulff et al., 2023).

Two theory streams explain diverging implications of GenAI usage on students' exam performance. *Cognitive load theory* (cf. Sweller, 1988; van Merriënboer & Sweller, 2005) suggests learning can be enhanced by reducing extraneous cognitive load. Thus, GenAI tools can be used as cognitive aids for simplifying or scaffolding information processing, supporting learning efficiency and allowing students to retain information better, especially when dealing with complex material. In contrast, the *constructivist theory of learning* (cf. Bada & Olusegun, 2015) highlights the importance of being actively involved in the learning process to achieve deeper understanding and build knowledge. By using GenAI, students may bypass this essential cognitive process of comprehension, analysis, and summarization.

Looking at the empirical research of GenAI usage, literature provides evidence for both theories. Studies have found that GenAI enhances students learning, measuring learning speed (Möller et al., 2024), self-reported performance (Shahzad et al., 2024), and increasing less tangible factors like motivation (Gao et al., 2024). Other research reports detrimental effects, such as GenAI leading to reliance on easy information access (Bastani et al., 2024) or superficial learning (Rasul et al., 2023). There are studies across the board identifying various ways GenAI can support or hinder student learning. However, the overall effect on students' exam performance remains unclear. We seek to bridge this gap by addressing the research question: *What is the effect of students' GenAI usage on their exam scores?*

To do so, we collect a sample of student data and empirically test the effect of GenAI usage on exam performance. We employ a fixed effects regression controlling for numerous factors affecting the exam score. Because students' use of GenAI for writing case study essays is not directly observable, we use ZeroGPT, a renowned GenAI detection system. In additional analyses, we utilize an identification strategy with exam retakers to further examine the causal effect and disaggregate the effect by considering students' learning potential. Our results show that students who use GenAI score significantly lower on exams. The negative effect is particularly large for students with high learning potential, indicating that GenAI use affects exam performance by impeding users' learning progress. Our finding holds after including several robustness checks. We can also rule out that our results are driven by lower scoring students having a higher tendency to use GenAI by controlling for possible confounding factors such as engagement and prior knowledge.

We contribute to existing research in several ways. First, we extend the information systems (IS) research on GenAI by examining its implications in higher education. A number of IS studies emphasize the technical capabilities and power of GenAI and its disruptive potential (e.g., Ahangama, 2023; Lee, 2023). Second, we contribute to higher education literature by investigating the effect of GenAI on exam scores. Educational institutions and researchers have discussed the costs and benefits of GenAI extensively, ultimately asking whether tools such as ChatGPT should be banned in higher education (Chhina et al., 2023; Kishore et al., 2023; Van Slyke et al., 2023). We not only provide evidence on the implications of GenAI for exam performance, but are also able to address effects on different user groups. Third, by identifying GenAI usage by detection tools, we extend research on GenAI detectors, discuss various approaches to identify AI-generated content, and document which GenAI detectors provide trustworthy results.

The paper is structured as follows: Section 2 covers the conceptual basics of GenAI and large language models (LLMs) and summarizes the relevant literature. In Section 3, we describe our multivariate model, discuss AI detectors, and provide details on our sample and descriptive statistics. Section 4 presents the empirical results, robustness checks, and additional analyses, which are discussed in detail in Section 5. In Section 6, the paper concludes with a summary, an examination of the study's limitations and an outlook for further research.

# 2 Conceptual Basics and Related Work

## 2.1 Generative AI and Large Language Models

GenAI refers to machine learning techniques (e.g., neural networks) to create seemingly novel and meaningful data instances or artifacts based on patterns and relationships in training data (Feuerriegel et al., 2024; Tao et al., 2023). These artifacts appear in various forms such as text, images, sound, and video (Alavi et al., 2024). LLMs are a subset of GenAI models capable of processing and creating natural language by applying learning technologies to extensive datasets (Lee, 2023). They can comprehend context and create textual data outputs similar to human language without requiring specific input formats (Teubner et al., 2023; von Brackel-Schmidt et al., 2023; Wilson et al., 2023).

GenAI constitutes the larger technological infrastructure required for the practical implementation of LLMs, including the actual model and user-facing components, their modality, and corresponding data processing (Feuerriegel et al., 2024). Such implementation enables users to enter input data and instructions conditioning the LLM, which is referred to as *prompting* (Feuerriegel et al., 2024; von Brackel-Schmidt et al., 2023). With the emergence of conversational LLMs (e.g., models with a chat-based interface), prompting shifted from one-off inputs toward multi-step interactions (von Brackel-Schmidt et al., 2023). These GenAI models are capable of completing various tasks, such as developing creative ideas, software coding, and textual content creation with high accuracy in grammar and wording (Yuan & Chen, 2023). These capabilities render GenAI particularly interesting for knowledge work as in academia and higher education (Benbya et al., 2024; Yuan & Chen, 2023).

## 2.2 Literature Review and Research Question

A considerable body of literature has rapidly emerged discussing how GenAI influences learning behaviour and success. GenAI seems to offer several benefits for learning, that could potentially increase exam performance. For example, Khatib and Mattalo (2024) and Fauzi et al. (2023) report that GenAI-based chatbots help students by providing answers to unclear questions. However both studies do not test for a tangible effect on exam scores. Möller et al. (2024) find a 27 % increase in students' learning speed when using GenAI as a chatbot. Similarly, Shahzad et al. (2024) document a significant increase in self-reported learning performance among students using GenAI-based technologies for learning. Other empirical studies find more latent benefits. Cotton et al. (2023) document increased student engagement and collaboration, while other studies mention higher motivation when studying is supported by ChatGPT (Fauzi et al., 2023; Gao et al., 2024). Further studies claim that GenAI is particularly helpful for disadvantaged or less privileged students, such as non-native speakers or those with communication disabilities. Cheon et al. (2024) and Tsekouras et al. (2024) report that providing plain language explanations and reducing complexity of teaching materials helps those students keeping up, however do not test for tangible performance effects.

These positive findings align with the *cognitive load theory* (cf. Sweller, 1988; van Merriënboer & Sweller, 2005). It assumes that inherent limitations of learning result from working memory load, which dichotomises into intrinsic and extraneous cognitive load. Intrinsic cognitive load is determined by the interaction between the learning material and the expertise of the learner. In contrast, extraneous cognitive load is not necessary for learning and can be reduced by instructional interventions (van Merriënboer & Sweller, 2005). GenAI tools can serve as cognitive aids (e.g., by providing summaries), thus reducing extraneous cognitive load when leaning. Unburdening human working memory may increase learning capacity ultimately enhancing exam performance.

In contrast, there are also studies reporting negative consequences of GenAI usage on learning. Easy access to answers without the need for close and detailed engagement with materials may lead to superficial learning (Rasul et al., 2023), which could hinder students' ability to deeply understand learning materials (Crawford et al., 2023). Abbas et al. (2024) provide empirical evidence for this in a structural equation model by showing that GenAI usage leads to procrastination and memory loss, ultimately correlating with a lower GPA. Similarly, Bastani et al. (2024) report initial improvements in scores when students are forced to use GenAI for an experimental exam. However, when GenAI assistance is taken away in a subsequent experiment, students scored lower than those who never had access. This suggests that students who study with GenAI will perform worse in exams as GenAI will not be available then. Moreover, the ability of GenAI to facilitate writing texts, essays, and note taking negatively affects students' learning (Lund et al., 2023). According to Milano et al. (2023) it diminishes the effort involved in crafting well-written and argued texts—effort that helps in understanding course materials and which has a positive influence on exam performance.

These findings rather align with the *constructivist theory of learning* (cf. Bada & Olusegun, 2015). It claims that being actively involved in the learning process archives deeper understanding. When students immerse themselves in their subjects, they are more likely to experience an *eureka* moment that enhance comprehension. By using GenAI for essay writing, students may bypass this essential cognitive process of comprehension, analysis, and summarization. This might similarly occur when GenAI is used to study for exams. If for instance GenAI is used to simplify or explain complex topics, students might use it as a shortcut that makes learning seem easier, but which actually prevents them from going through the process of understanding and learning on a deeper level.

In summary, both theory and empirical research has identified various ways in which GenAI can support or hinder learning, presenting a mixed picture of its impact on students' performance. The effect of GenAI on students' actual exam score has not yet been examined. Ultimately, assessment in higher education rests on exam performance; hence, a comprehensive understanding of GenAI tools in higher education requires investigation of this tangible effect, which so far has not been thoroughly explored. Our study seeks to examine GenAI's overall impact on exam performance. To do so, we analyse the effect of students' GenAI use on their exam scores. Focusing on exam scores provides a measure that encapsulates the individual effects of GenAI on learning, offering a comprehensive view of its impact on student performance.

## 3 Data and Methodology

### 3.1 Multivariate Model and Approach

To answer our research question, we utilize the educational setting of our institution's first-year introductory accounting class. To detect GenAI users among the cohort, we rely on case study essays our students submitted during the semester. The case study concerns a knowledge transfer exercise for students to immerse themselves in the course material and enhance comprehension. We identify GenAI-written texts using ZeroGPT, a popular and frequently used online GenAI detector. This measure of GenAI usage allows us to empirically assess its impacts on exam scores using a fixed effects ordinary least square (OLS) regression. In line with related research (e.g., Chiu et al., 2023), we control for various factors that have been shown to affect exam scores. The full OLS model reads as follows:

$$Exam\ Score_i = \beta_0 + \boldsymbol{\beta_1}\ \boldsymbol{GenAI\ User_i} + \beta_2\ A-Level\ Grade_i + \beta_3\ Attempt_i$$
$$+ \beta_4\ Attendance_i + \beta_5\ Vocational\ Training_i + \beta_6\ Voluntary\ Service_i \quad (1)$$
$$+ \beta_7\ Female_i + \beta_8\ LinkedIn\ User_i + Course\ of\ Study\ FE + \varepsilon_i$$

Our dependent variable is *Exam Score*, which is a continuous measure indicating the percentage of points a student achieved in the final exam. While the minimum is zero, the actual (achievable) maximum is 96.67 (100). The variable of interest is *GenAI User*, an indicator variable taking the value one if a student uses GenAI for studying and for producing work that the instructor intended to be written without such assistance, and zero otherwise. Based on the indicated probability of our GenAI detector, we classify students as *GenAI User* if the text is more likely to have been written by AI than not (percentage > 50 %).[1]

To eliminate potential confounding effects biasing our inference, we use established determinants of exam performance as control variables. First, we control for academic preparedness and achievements prior to higher education. We include *A-Level Grade* as a common predictor for exam scores (e.g., Lento, 2018; Massoudi et al., 2017). We also include two dummy variables indicating completion of *Vocational Training* or *Voluntary Service* as indications of maturity and experience (Guney, 2009; Hartnett et al., 2004; Voshaar, Wecks, et al., 2023). Second, we control for academic behaviour as it directly influences exam performance by including session *Attendance* and the number of *Attempts* at taking the final exam (Cheng & Ding, 2021; Massoudi et al.,

[1] Alternatively, and to rule out potential biases, we use higher (0.6) and lower (0.4) thresholds in ZeroGPT, use the detection score as continuous variable and other GenAI detection tools to define our variable of interest (see our robustness checks).

2017; Voshaar, Knipp, et al., 2023; Voshaar et al., 2024). Third, previous studies have found correlations between exam performance and gender as well as course of study (Aldamen et al., 2015; Hu et al., 2023; Wecks et al., 2023). We include *Course of study*-fixed effects and a dummy variable indicating students' gender (*Female*). In addition, we introduce a novel control variable by adding a dummy indicating whether a student is a *LinkedIn User*. LinkedIn usage has been found to be correlated with exam performance (Paul et al., 2012). GenAI acceptance is driven by personal innovativeness and openness to technology (Strzelecki, 2023), which is also correlated with (new) social media usage, such as LinkedIn (e.g., Wijesundara & Xixiang, 2018). Online Appendix A comprises a table with all variables including definition, rationale, and references (https://tinyurl.com/zjehfa3n).

### 3.2 Generative AI Detection Systems

To differentiate between GenAI users and non-users, we utilize the GenAI detection tool ZeroGPT. With the widespread availability of GenAI, also the dissemination of detection tools has accelerated (e.g., Dalalah & Dalalah, 2023). Detection systems split the inputted text into individual tokens and predict the probability that a specific token will be followed by the subsequent sequence (Crothers et al., 2023). The detector also analyses a text's perplexity, which refers to the use of random elements and idiosyncrasies typical of human writing and speech (Walters, 2023). If the detector identifies high predictability and low perplexity, it is probable that the text is AI-written and is recognized as such.

Based on research and test runs, we opted to use ZeroGPT (https://www.zerogpt.com/) for multiple reasons. First, ZeroGPT works with German texts. Second, research found ZeroGPT to be among the best detector tools, with consistent and accurate GenAI detections (Walters, 2023; Weber-Wulff et al., 2023). Third, ZeroGPT has been shown to perform well at avoiding both false positives and negatives in GenAI detection (Walters, 2023). Finally, ZeroGPT can identify content generated by all state-of-the-art GenAI models, such as ChatGPT, Gemini, and LLaMA. When a text is inputted, ZeroGPT uses machine learning algorithms and natural language processing to analyse it and identify common GenAI patterns (Alhijawi et al., 2024; ZeroGPT, 2024). The output is the proportion of tokens estimated to be AI-generated, which is more detailed than the binary outputs of several other detectors (e.g., Copyleaks).

### 3.3 Sample Selection and Descriptive Statistics

Our study is based on a broad sample of business, economics, and management students taking an introductory financial accounting course at a German university in the winter term 2023/2024. To obtain the required data for our analysis, we have drawn on several data sources. First, using an online survey at the beginning of the semester, we collected data on student characteristics that might influence exam scores. Second, we retained the students' case study essays throughout the semester, and processed and analysed them in terms of GenAI usage after the final exam. Third, we obtained the final exam scores from the central examination office to evaluate the impact on exam scores.

Starting with 572 students who participated in the survey, we first excluded students who did not hand in an essay ($N$ = 243). Additionally, we excluded those students who did not take the final exam ($N$ = 127). Finally, we dropped the observations with missing data for the control variables ($N$ = 9). This leaves us with a final sample of 193 students. Given the 502 students in the final exam, our sample accounts for about 38 % of the underlying population and can thus be considered representative.

Table 1 reports the descriptive statistics of the student characteristics for the full sample.[2] The mean exam score is 45.39, indicating that our sample students on average fall below the 50 % threshold. This reflects the high failure rates commonly observed among higher education introductory accounting classes (Prinsloo et al., 2010; Sanders & Willis, 2009). The binary variable of interest *GenAI User* has a mean of 0.306, indicating that 30.6 % of the students in our sample (i.e., 59 students) are identified as GenAI users by ZeroGPT. The mean value of *ZeroGPT* indicates that on average 35.4 % of students' texts are flagged as AI-generated. The average student in our sample has taken the final exam for the first time (mean *Attempt* = 1.425) and has attended fewer than half of the offered tutorials (mean *Attendance* = 0.447). A rather small subset of 19.2 % of the students has a LinkedIn profile.

[Insert Table 1 about here]

We also test for the differences between the variables divided into GenAI user and non-user group. The average *GenAI User* achieves 9.027 ($p$-value < 0.01) fewer exam points. Besides the exam score, there are no statistically significant differences between the user and non-user group, with the exception of a borderline significant difference in A-Level grade and attempt.[3] This gives an initial indication of poorer exam performance among GenAI users compared to non-users. However, because not only GenAI usage but also student characteristics may influence exam scores, the association of GenAI usage and exam scores calls for multivariate examination taking control variables into consideration.

---

[2] Pearson correlations are reported in Online Appendix B (https://tinyurl.com/zjehfa3n) and do not show any indication of multicollinearity issues, as the highest absolute value is 0.331.

[3] Online Appendix C shows all univariate differences (https://tinyurl.com/zjehfa3n).

# 4 Impact of Generative AI Usage on Exam Performance

## 4.1 Main Results

Table 2 shows the multivariate regression results hierarchically. In column (1), we regress the exam score on *GenAI Usage* along with common control variables. The coefficient of the variable of interest is statistically significant and negative. We add course of study as fixed effect in column (2). Column (3) additionally includes *LinkedIn User* as another control variable.[4] The coefficients of the control variables are consistent with the literature. For the variable of interest (*GenAI User*), we continue to find a significantly negative coefficient. The results suggest that students using GenAI score 6.71 points lower in the final exam, which is substantial, as the mean student scores 45.39 points. Thus, on average, the scores of GenAI users are about 15 % lower than that of the mean non-user. Given that the passing threshold is at 41 points, GenAI use can tip the scales toward failing the exam—at least statistically. The empirical evidence provides a clear picture of a negative impact of GenAI usage on exam scores.

[Insert Table 2 about here]

## 4.2 Robustness Checks

We conduct a set of robustness checks to ensure the reliability and validity of our results. Our initial step involves scrutinizing the robustness of the *GenAI User* variable. In our main analysis, we identify GenAI users based on a threshold of over 50 % in the written case study essays, as determined by ZeroGPT. This approach yields 30.6 % of our sample as GenAI users. To test whether this share is realistic, we conduct an anonymous survey among all students in our sample (see Online Appendix D: https://tinyurl.com/zjehfa3n). Among the 30 survey responses (15.5 % of the sample), 30.0 % state that they used GenAI tools for the written case study, aligning well with our measured value. Identifying about a third of our population as GenAI users is also consistent with findings reported elsewhere in the literature. von Garrel and Mayer (2023) conducted a representative survey among German university students and find that 34.8 % of them report using AI-based tools for studying occasionally, frequently, or very often. Considering the variation in reported usage rates across different studies and online reports, we proceed to test alternative thresholds for the detection tool. Adjusting the threshold to 0.6 reduces GenAI users to 20.2 %, while a threshold of 0.4 increases them to 40.9%. The results in columns (1) and (2) of Table 3 with the adjusted thresholds remain robust and unchanged. We also use the detection score as a continuous independent variable in column (3), indicating that more intense usage decreases exam scores.

---

[4] The *link* test (Pregibon, 1980), a significant *F*-test, and the coefficient of determination (adjusted $R^2$) all indicate a well-fitted model. As the Breusch and Pagan (1979) and Cook and Weisberg (1983)-test detects no heteroscedasticity (*p*-value of 0.56), we refrain from using robust standard errors.

Additionally, we show that our findings are robust when using other detection tools. We repeat the analysis using an AI detector particularly designed for the German language, developed at the University of Applied Sciences Wedel (Tlok et al., 2023). According to the developer, this tool's outputs are probabilities and thus not directly comparable to other tools. As shown in Online Appendix E (https://tinyurl.com/zjehfa3n), this results in a distribution of outputs with many at 0 % probability and a uniform distribution across the remaining value range. A *more likely than not* classification would be impractical. Instead, we run a pre-test with 12 student seminar papers written before GenAI was available and modify them using ChatGPT 4, creating a paired sample of known GenAI and non-GenAI texts on the same topics as our main study. The German detector consistently shows values below 10 % for human-written and above 10 % for AI-generated texts.

Adopting this threshold for our robustness check, we identify 36.3 % of our sample as GenAI users, which is similar to the main analysis, our survey, and the values reported in previous literature (von Garrel & Mayer, 2023). Column (4) presents the results using the German detector. The *GenAI User* coefficient is now even more negative and significant, suggesting improved accuracy due to the detector's optimization for German texts.

[Insert Table 3 about here]

To address potential concerns about the opacity of AI detectors, we conduct a further robustness check by manually computing the propensity of GenAI usage. Literature reports that GenAI-generated texts typically exhibit lower readability, higher lexical richness, and a greater number of adjectives than human-written texts (Muñoz-Ortiz et al., 2023; Shah et al., 2023). GenAI texts tend to have more words per sentence and sentences per paragraph, contributing to lower readability scores (Deveci et al., 2023; Pehlivanoğlu et al., 2023). We employ the Gunning-Fog Index as a well-regarded measure of readability (Gunning, 1952). Lexical richness essentially refers to the ratio of unique words measurable by the metric Herdan's C (Herdan, 1960). As another metric, we consider the proportion of adjectives used (Markowitz et al., 2023). We conduct a principal component analysis to consolidate these three variables into a single vector.[5] This results in a factor variable ranging from 0 to 1, indicating GenAI markers. In column (5), we use this factor as variable of interest and find that the presence of lexical characteristics of GenAI-generated texts in a student's work correlates with lower exam scores, reinforcing the findings from previous analyses.

[5] The Kaiser-Meyer-Olkin measure of 0.657 and a significant Bartlett test of sphericity (*p*-value < 0.01) indicate that the variables are highly correlated and collectively measure the construct of a text being written by GenAI.

Beyond the robustness of our measure, we address potential endogeneity issues. Group-wise comparison between GenAI users and non-users indicates that GenAI users have a lower A-level grade and a higher number of attempts. Academic preparedness or experience might affect both exam performance and GenAI usage. Our approach already addresses potential endogeneity to some extent, as our control variables can capture such characteristics (Hill et al., 2021). To provide additional robustness, we use entropy-balancing (Hainmueller, 2012) to adjust the weights of control observations to ensure the means and variances of control variables are identical in the treatment and control group, minimizing selection bias. The results in column (6) underscore our main findings even when controlling for potential endogeneity.[6]

### 4.3 Additional Analyses

Our main results show a negative impact of GenAI usage on the exam score. To further explore the effect and the mechanism at work, we conduct additional analyses. We explore the effect for different levels of student engagement and cognitive abilities and measure how GenAI usage affects performance improvement when repeating the exam.

First, we apply an identification strategy to further analyse the causal effect. We identify and match all repeating students who did not pass the exam in the year before, when GenAI models were not yet available for student use (*Pre GenAI*).[7] This leaves us with a sample of matched observations before and after broad GenAI availability containing (*Pre*) *GenAI Users* (*N* = 15) and (*Pre*) *Non-Users* (*N* = 12). Figure 1 shows the distribution of exam scores for each group and the differences in mean exam scores between the groups and time periods (within group) along with their statistical significance. Due to the small sample size, we bootstrap the distributions around the mean.

---

[6] This approach primarily addresses observable sample selection bias, while this issue might also arise from omitted correlated variables (unobserved sample selection bias). However, our course of study-fixed effects mitigate this to some extent (Wooldridge, 1995) and the Ramsey (1969) *RESET* test indicates no omitted variables (*p*-value of 0.764). Additionally, potential instrumental variables related to technical affinity and engagement do not show any correlation with *GenAI User*, allowing us to discount endogeneity due to omitted correlated variables.

[7] The *Pre GenAI* semester ended in January 2023. The first publicly available GenAI model (ChatGPT 3) was launched only a few weeks before the exam but had very few users, difficult access, and extensive downtime at that early stage. Therefore, an effect on the exam performance can be ruled out, allowing us to consider this semester as a *Pre GenAI* period.

[Insert Figure 1 about here]

The *Pre GenAI* period solely consists of students who failed the exam. We observe a statistically significant improvement in exam score in the next attempt for both groups (*Post GenAI*).[8] However, the *GenAI User* group shows a substantially lower increase. While both subsamples perform equally well in their second attempt, those in the *GenAI User* group reach far more points than the *Non-User group* in the attempt before in the *Pre GenAI* period. In the attempt after GenAI was widely available, the *Non-User* group increases their exam scores to a greater extent than the *GenAI User* group. Consequently, we observe a learning-hindering for students using GenAI, as they improve far less.

In our second additional analysis, we perform split sample analyses using two measures of student capabilities and engagement. If the documented effect in our main results is indeed attributable to hindering learning, the effect should be stronger for students where individual learning and comprehending the content would otherwise have fallen on fertile ground. To approximate this characteristic, we use A-level grades and attendance at tutorials. While the first addresses academic preparedness, pre-university achievements, and cognitive abilities, the latter measures engagement. We conduct a median split for both measures to create two samples with low and high A-level grades and attendance to gain additional insights into the mechanism behind the effect. Table 4 shows the results of the additional split sample analysis. Columns (1) and (2) include the two regressions for the split samples by A-level grade, while attendance is used to split the sample in columns (3) and (4).

[8] An increased performance among repeating students aligns with related research attributing the effect to increased commitment (Martínez & Martinez, 1992; Voshaar, Knipp, et al., 2023; Wecks et al., 2023).

[Insert Table 4 about here]

For the split sample of students with good A-level grades (and therefore strong cognitive abilities demonstrated through considerable pre-university performance), we find the coefficient of *GenAI User* to be highly significant and negative (column (1)). While this aligns with our main results, the coefficient's magnitude is almost doubled compared to the one in column (3) of Table 2. In contrast, the coefficient is positive but insignificant for students with A-level grades below the median (column (2)). A similar picture emerges when using tutorial attendance (and hence engagement) for median split. While the students with higher attendance perform worse not only statistically significantly but to an educationally impactful extent when using GenAI, the effect is insignificant for low-attendance students (columns (3) and (4), respectively). This suggests that the negative effect of GenAI in the main analysis is primarily due to the effect on students with good A-level grades and high attendance.

We can conclude that the impact of GenAI use on exam performance varies depending on students' prior academic achievements and/or cognitive abilities as well as engagement during the semester. We find using GenAI to be detrimental to the exam scores of higher achieving and more engaged students. This confirms the learning-hindering mechanism as those students who would have been well equipped to understand the learning materials suffer particularly from the forgone opportunity to engage with the course content. When compared intra-group with other students with good prerequisites and who prepare for and write the essays themselves, the disadvantage of (over-)reliance on GenAI is even more glaring.

## 5 Discussion and Implications

Our main results show that GenAI usage negatively affects students' exam scores. While the literature has found many aspects of GenAI that can have a positive or negative influence on exam performance, it is unclear whether the benefits or the downsides prevail. We observe clear evidence of a negative overall effect on exam performance in the case that students use GenAI for case study writing. Positive aspects such as summarizing information (Pavlik, 2023), increasing study motivation (Fauzi et al., 2023), or providing plain-language explanations (Sullivan et al., 2023) may still occur. However, our results show that these are overshadowed by the negative effects that have been described in previous studies (Crawford et al., 2023; Rasul et al., 2023).

Looking at the causality, we can rule out that low exam scores lead to GenAI usage directly, as we measure the usage before the exam was taken. Yet, there may be characteristics associated with both exam score and propensity to use GenAI. Low engagement or poor prior knowledge lead to inferior exam scores and might also drive the use of GenAI, for example as a shortcut or assistance to keep up. However, our empirical model controls for these factors by including control variables such as in-class attendance, A-levels grade, and voluntary service that are related to engagement or prior knowledge. Thus, we conclude that we are not only measuring an association, but an effect of GenAI usage on exam scores.

We explore our effect in greater depth, leveraging an identification strategy in a sample of repeating students and find that students opting for GenAI usage do not exhibit an increase in their exam scores similar to that of their peers not using GenAI. We ascribe this result primarily to a learning-hindering mechanism. When students let GenAI write an essay on a complex and challenging topic, instead of exploring, grappling with, and mastering the content and then writing the essay themselves, students waste the opportunity to learn and to experience the inherent rewards of figuring things out. Research such as Milano et al. (2023) warn of this, stating that GenAI's role in facilitating academic writing is dependable to the point of negatively impacting students' learning. Similarly, the ready availability of quick answers may reduce the intensity of students' engagement with the subject matter and ultimately deter their learning, as argued by studies such as Cotton et al. (2023) or Sallam et al. (2023).

The results of our split sample regressions further support the learning-hindering mechanism, as students with high learning potential are especially impacted by GenAI usage. The significantly negative effect for the students in our sample with good A-level grades or high attendance—which can be reasonably equated to higher levels of skill or commitment—indicates that these students have more to lose when they do not immerse themselves in the subject matter. And indeed, we find no effect for the opposite group, who are less predisposed to assimilate knowledge due to lower attendance or cognitive ability. We can however document the impact of GenAI usage for these students when looking at the results of the identification strategy analysis. This subsample solely consists of repeating students with attendance and A-Level grades that are below average. While we document a significant performance increase in the next exam attempt, GenAI usage hampers this improvement. Thus, the negative GenAI influence becomes evident when there is considerable learning potential, providing further support for the learning-hindering mechanism.

These findings align with the constructivist theory of learning (cf. Bada & Olusegun, 2015). Writing a case study essay is valuable not merely as an end in itself but also as an instrument to assist students in engaging with, exploring, and understanding subject-related content. However, when students utilize GenAI tools as a shortcut to avoid workload, they also bypass the cognitive process of comprehension, analysis, and summarization. This prevents students from retaining learning materials and understanding them on a deeper level. This detrimental use of GenAI prevents full exploitation of its potential in the learning process, as anticipated of cognitive load theory. The negative effect of GenAI usage we observe in this study suggests that students are not leveraging GenAI tools as cognitive aids to reduce extraneous load but rather as a shortcut avoiding full immersion with the learning material.

Our findings have important implications for students, educators, and educational institutions. For students, the results suggest a cautious approach to using GenAI for learning. While GenAI may appear to ease the learning process, it can adversely affect exam performance. Students should be mindful of the potential drawbacks and consider integrating GenAI as a supplementary tool rather than a primary resource for grappling with complex topics. Educators likewise need to take students' use of GenAI into account when designing curricula. It is essential to provide tasks and learning materials that promote deep learning and minimize the potential for GenAI to diminish engagement with the subject matter. Strategies could include incorporating in-class discussions, handwritten assignments, and other methods that encourage active learning and critical thinking. Lastly, this study offers valuable insights for educational institutions regarding GenAI policies in higher education. Although our results point to negative implications of GenAI use on student performance, we do not advocate for outright bans. As with many revolutionary information systems, there are both positive and negative aspects of GenAI. The negative results of this study may rather show that higher education does not yet harness the full potential of GenAI. Educational institutions should guide educators on how to instruct students in the proper use of GenAI and develop policies that mitigate its negative effects while amplifying its benefits. Similar to the disruptions of higher education caused by calculators and the internet, banning is not a practical solution. Students will inevitably encounter GenAI outside the university setting and must learn to use it effectively rather than confining themselves to getting by without it.

## 6 Conclusion, Limitations, and Future Research

The present study contributes to the rich debate on how GenAI will affect learning in (higher) education by evaluating the tangible effects of GenAI usage on exam performance. We address an important research gap, as performance effects have not yet been examined but are nevertheless crucial when discussing how to adapt education to the age of GenAI. Our findings reveal that using GenAI for writing essays significantly decreases exam scores. The additional analysis offers nuanced insights by documenting a learning-hindering mechanism through with GenAI usage negatively affects exam scores. Our study thus has implications for students, educators, and institutions.

Some of the limitations of our study warrant further attention. First, using GenAI detector tools to identify GenAI users among our students comes with the risk of inaccurate detection results. Not identifying all GenAI users (or too many) would affect our inference. Having said that, we improve the robustness of our results by using different approaches to detect GenAI usage. We also find the share of detected GenAI users to align with the numbers found in related research (von Garrel & Mayer, 2023) and through our own anonymous survey. Second, while our results may lack generalizability as our study is limited to a financial accounting class at one German university, several characteristics of the course suggest broader applicability. As an interdisciplinary course, it includes students from various fields such as business, economics, engineering, and computer science. Our sample also spans students at different stages of their academic careers, from first-year students to those nearing graduation. The course format—a large lecture with a final exam—is typical of many university classes across continental Europe, making the setting comparable to similar academic contexts. As a final limitation, our study does not account for usage behaviour and hence usage intensity. Different intensities of use might come with different impacts on exam performance. Similarly, our findings only apply for case study writing as part of the broader learning experience and in the context of a final written exam. However, GenAI can be used for various learning activities, such as a learning companion, personalized tutor, and to summarize complex content in a student-friendly manner. Also, academic performance covers a wide range of outcomes not being limited to written exams. Our findings do not necessarily hold in the context of other assessment types.

Therefore, these limitations can serve as a springboard for future research, which could examine how GenAI affects exam performance when used for other learning purposes than case study writing. We additionally call for research that complements our findings by exploring the impact of GenAI usage on different learning outcomes. Future studies could examine the effect on presentation-based performance assessments, oral exams, and term papers and theses. Future research could finally consider how students use GenAI to help them write case study texts or explore their usage behaviour in more depth. Exploring the threats and opportunities from this perspective would help institutions position themselves in discussing whether using GenAI tools should be banned, tolerated, or taught.

## Figures and Tables

<table>
<tr><th>Group</th><th>Pre GenAI</th><th>Post GenAI</th><th>|Mean Diff|</th></tr>
<tr><td>GenAI User<br>(N = 15)</td><td>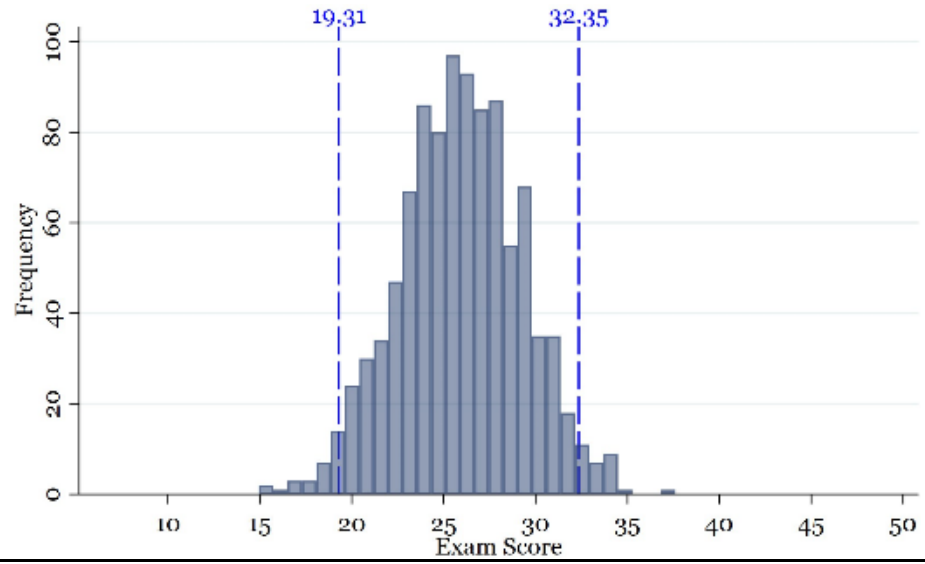
</td><td>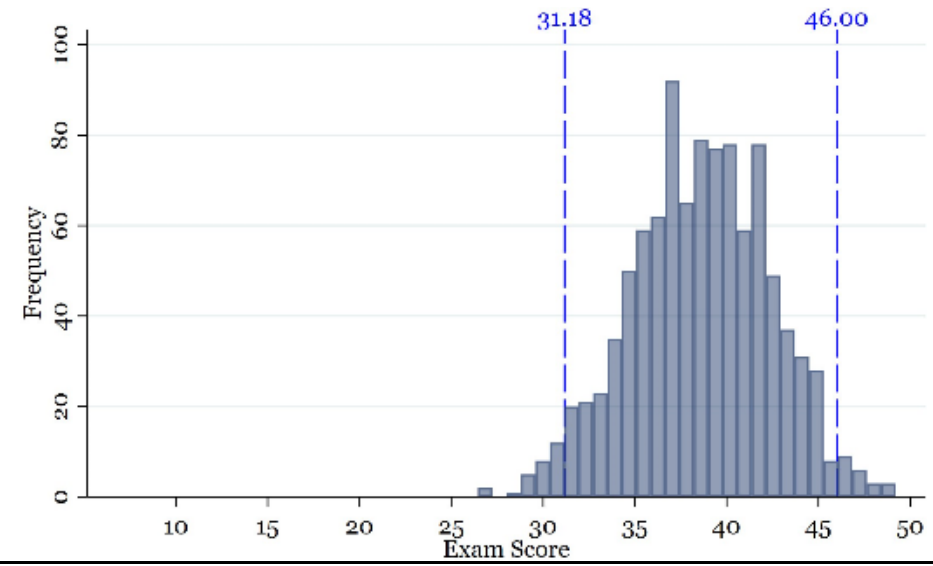
</td><td>12.76**</td></tr>
<tr><td>Non-User<br>(N = 12)</td><td>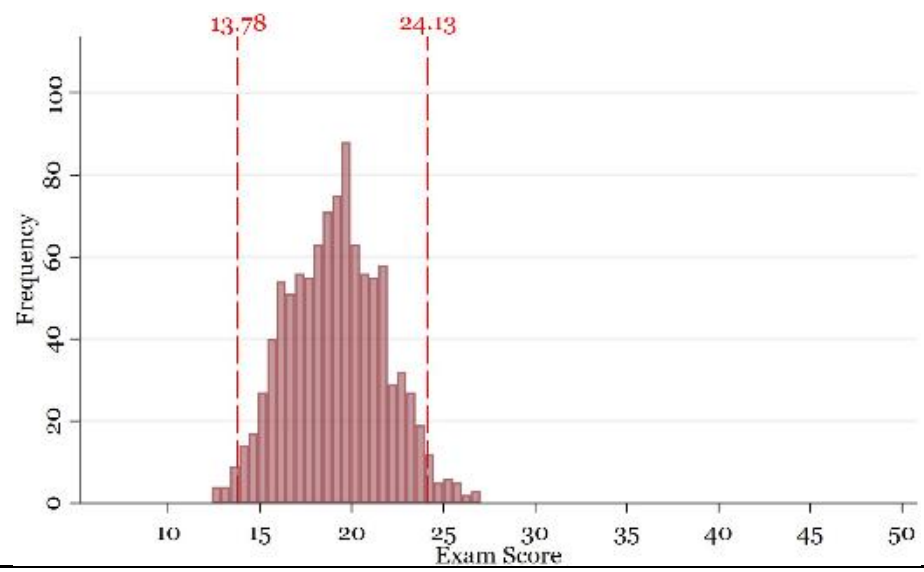
</td><td>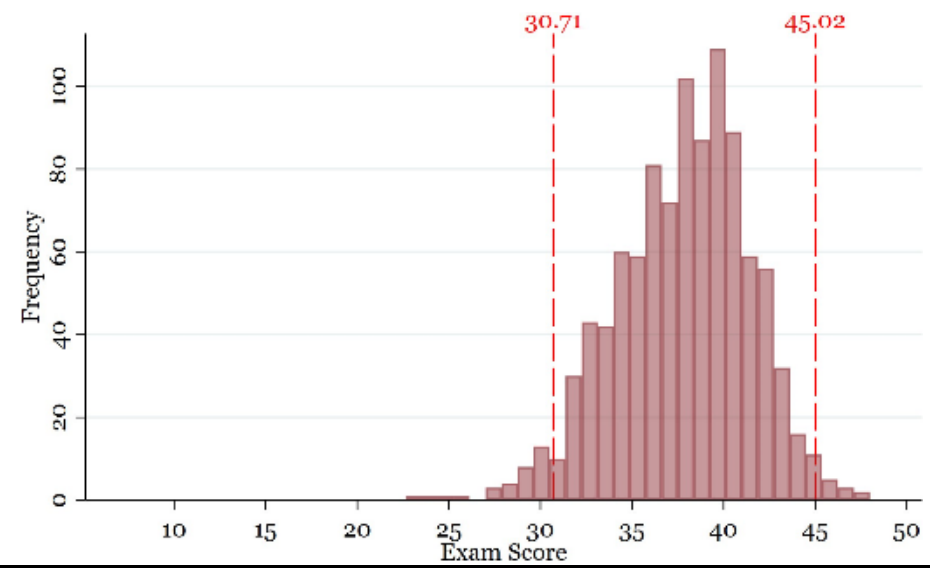
</td><td>18.91***</td></tr>
<tr><td>|Mean Diff|</td><td>6.86</td><td>0.72</td><td></td></tr>
<tr><td colspan="4">Figure 1 shows the bootstrap distributions (1,000 replications) and 95 % confidence interval of exam scores from the four groups of (Pre) GenAI User and Non-User before and after the public release of GenAI. Also, the mean differences between GenAI user and non-user performance are reported as absolute values and tested by a paired sample (two-sample) t-test for within- (between-)group difference. ***, **, * indicate statistical significance at the 1 %, 5 %, and 10 % level, respectively (two-tailed).</td></tr>
</table>

**Figure 1 Results of the Identification Strategy**

| Student Data | *N* | *Mean* | *Median* | *SD* | *P25* | *P75* |
|---|---|---|---|---|---|---|
| *Exam Score* | 193 | 45.39 | 45.83 | 22.23 | 27.50 | 61.11 |
| *GenAI User* | 193 | 0.306 | | 0.462 | | |
| *ZeroGPT* | 193 | 0.354 | 0.296 | 0.277 | 0.119 | 0.556 |
| *A-Level Grade* | 193 | 2.290 | 2.200 | 0.607 | 1.800 | 2.700 |
| *Attempt* | 193 | 1.425 | 1 | 1.223 | 1 | 1 |
| *Attendance (relative)* | 193 | 0.447 | 0.444 | 0.322 | 0.111 | 0.778 |
| *Vocational Training* | 193 | 0.135 | | 0.342 | | |
| *Voluntary Service* | 193 | 0.363 | | 0.482 | | |
| *Female* | 193 | 0.472 | | 0.500 | | |
| *LinkedIn User* | 193 | 0.192 | | 0.395 | | |

Table 1 shows the descriptive statistics. For binary variables, only means and standard deviations are presented.

**Table 1 Descriptive Statistics**

| Variables | (1) | (2) | (3) |
| --- | --- | --- | --- |
| ***GenAI User*** | **-6.87 **** | **-6.32 **** | **-6.71 **** |
| | **(-2.26)** | **(-2.01)** | **(-2.17)** |
| *A-Level Grade* | 12.19 *** | 12.04 *** | 11.42 *** |
| | (5.26) | (4.98) | (4.79) |
| *Attempt* | 1.05 | 0.93 | 0.76 |
| | (0.87) | (0.77) | (0.63) |
| *Attendance (relative)* | 19.71 *** | 17.99 *** | 18.30 *** |
| | (4.31) | (3.73) | (3.86) |
| *Vocational Training* | 9.76 ** | 10.54 ** | 10.80 ** |
| | (2.32) | (2.41) | (2.52) |
| *Voluntary Service* | 2.15 | 2.20 | 1.76 |
| | (0.71) | (0.72) | (0.59) |
| *Female* | -8.61 *** | -9.54 *** | -10.16 *** |
| | (-3.10) | (-3.23) | (-3.50) |
| *LinkedIn User* | | | 9.60 *** |
| | | | (2.75) |
| *Constant* | Included | Included | Included |
| *Course of Study-Fixed Effects* | - | Included | Included |
| *N* | 193 | 193 | 193 |
| *Adj.* $R^2$ | 0.28 | 0.28 | 0.30 |

Table 2 presents the regression results with exam score as dependent variable. In column (1), the independent variables are the GenAI usage dummy variable and control variables commonly found in literature. Column (2) adds students' course of study as fixed effects. Column (3) then depicts our main results, including another control variable. Bold font indicates the variable of interest. ***, **, * indicate statistical significance at the 1 %, 5 %, and 10 % level (two-tailed), respectively. *t*-values are presented in parentheses. All variables are defined in Online Appendix A (https://tinyurl.com/zjehfa3n).

**Table 2 Regression Results on the Impact of GenAI Usage on Exam Score**

| Variables | (1) Threshold 0.4 | (2) Threshold 0.6 | (3) Continuous Detection Score | (4) German Detector | (5) Manual Computation | (6) Balanced Sample |
|---|---|---|---|---|---|---|
| ***GenAI User*** | **-7.10 **** | **-7.17 **** | **-14.67 ***** | **-8.53 ***** | **-2.66 ***** | **-6.51 **** |
| | **(-2.43)** | **(-2.02)** | **(-2.82)** | **(-2.90)** | **(-2.70)** | **(-2.07)** |
| *A-Level Grade* | 11.64 *** | 11.42 *** | 11.46 *** | 11.21 *** | 11.62 *** | 8.88 *** |
| | (4.92) | (4.78) | (4.81) | (4.75) | (4.87) | (3.03) |
| *Attempt* | 0.89 | 0.84 | 0.96 | 0.94 | 0.79 | 2.05 ** |
| | (0.74) | (0.69) | (0.79) | (0.79) | (0.66) | (1.98) |
| *Attendance (relative)* | 17.70 *** | 18.38 *** | 14.85 *** | 16.81 *** | 15.55 *** | 17.75 *** |
| | (3.75) | (3.87) | (3.78) | (3.58) | (3.25) | (2.95) |
| *Vocational Training* | 10.37 ** | 11.25 *** | 10.92 ** | 10.93 ** | 12.09 *** | 8.75 |
| | (2.41) | (2.63) | (2.57) | (2.58) | (2.86) | (1.65) |
| *Voluntary Service* | 1.81 | 2.07 | 2.60 | 1.57 | 1.24 | 1.52 |
| | (0.60) | (0.69) | (0.87) | (0.53) | (0.41) | (0.45) |
| *Female* | -10.14 *** | -9.29 *** | -9.40 *** | -10.20 *** | -10.18 *** | -10.49 *** |
| | (-3.50) | (-3.18) | (-3.24) | (-3.54) | (-3.50) | (-3.22) |
| *LinkedIn User* | 10.75 *** | 9.25 *** | 9.34 *** | 10.26 *** | 7.89 ** | 10.53 *** |
| | (3.05) | (2.65) | (6.66) | (2.96) | (2.25) | (2.82) |
| *Constant* | Included | Included | Included | Included | Included | Included |
| *Course of Study-FE* | Included | Included | Included | Included | Included | Included |
| *N* | 193 | 193 | 193 | 193 | 193 | 193 |
| *Adj. $R^2$* | 0.31 | 0.30 | 0.41 | 0.32 | 0.31 | 0.20 |

Table 3 presents the results of the robustness checks. In columns (1) and (2), we reduced (> 0.4) or increased (> 0.6) the threshold of the AI detector value to be classified in the *GenAI User* group. In column (3) we include the continuous detection score from ZeroGPT as our independent variable *GenAI User*. Column (4) uses alternative AI detectors. Column (5) includes a manual computed score that represents AI detection. In column (6), we again present our main results but with an entropy-balanced sample. Bold font indicates the variable of interest. ***, **, * indicate statistical significance at the 1%, 5%, and 10% level (two-tailed), respectively. *t*-values are presented in parentheses. All variables are defined in Appendix A (https://tinyurl.com/zjehfa3n).

**Table 3 Results of Robustness Checks**

| Variables | (1) Higher A-Level Grade | (2) Lower A-Level Grade | (3) Higher Attendance | (4) Lower Attendance |
|---|---|---|---|---|
| ***GenAI User*** | -12.27 *** | 2.09 | -11.90 *** | -2.92 |
| | (-2.73) | (0.46) | (-2.74) | (-0.64) |
| *A-Level Grade* | 11.96 ** | 24.13 *** | 11.58 *** | 12.64 *** |
| | (2.37) | (3.25) | (3.60) | (3.31) |
| *Attempt* | -2.03 | 1.21 | -1.78 | 2.27 |
| | (-0.87) | (0.87) | (-0.71) | (1.44) |
| *Attendance (relative)* | 17.18 ** | 12.07 * | 18.62 | 34.93 * |
| | (2.24) | (1.89) | (1.65) | (1.83) |
| *Vocational Training* | 4.41 | 24.40 *** | 10.33 * | 11.93 * |
| | (0.76) | (3.80) | (1.77) | (1.76) |
| *Voluntary Service* | -1.17 | 3.97 | -2.80 | 8.47 * |
| | (-0.27) | (0.94) | (-0.67) | (1.83) |
| *Female* | -9.65 ** | -9.44 ** | -11.07 *** | -8.22 * |
| | (-2.25) | (-2.30) | (-2.76) | (-1.71) |
| *LinkedIn User* | 9.87 ** | 6.00 | 16.44 *** | 0.81 |
| | (2.12) | (1.12) | (3.37) | (0.15) |
| *Constant* | Included | Included | Included | Included |
| *Course of Study-FE* | Included | Included | Included | Included |
| *N* | 103 | 90 | 104 | 89 |
| *Adj. $R^2$* | 0.27 | 0.28 | 0.30 | 0.12 |

Table 4 presents the regression results using split samples. In columns (1) and (2), we repeat our main regression analysis on a restricted sample only containing students with above- (below-)median *A-Level Grade*. Columns (3) and (4) present the main regression separately for students with above- and below-median attendance. Bold font indicates the variable of interest. ***, **, * indicate statistical significance at the 1%, 5%, and 10% level (two-tailed), respectively. *t*-values are presented in parentheses. All variables are defined in Appendix A (https://tinyurl.com/zje-hfa3n).

**Table 4 Results of Split Sample Regressions**